\documentclass[12pt]{iopart}
\usepackage{graphicx}
\usepackage{epsfig}
\expandafter\let\csname equation*\endcsname\relax
\expandafter\let\csname endequation*\endcsname\relax
\usepackage{amsmath}
\usepackage{iopams}

\newcommand{\bra}[1]{\ensuremath{\langle{#1}|\,}}
\newcommand{\ket}[1]{\ensuremath{\,|{#1}\rangle}}
\newcommand{\braket}[1]{\ensuremath{\langle{#1}\rangle}}

\begin{document}

\title{Nonequilibrium perturbation theory in Liouville-Fock space  for inelastic electron transport}

\author{Alan A. Dzhioev\footnote{On leave of absence from Bogoliubov Laboratory of Theoretical Physics, Joint Institute for Nuclear Research,  RU-141980 Dubna, Russia},
        D. S. Kosov}
\address{Department of Physics, Universit\'e Libre de Bruxelles, Campus Plaine, CP 231, Blvd du Triomphe, B-1050 Brussels, Belgium }
\ead{adzhioev@ulb.ac.be}


\begin{abstract}
We use   superoperator representation  of quantum kinetic equation to develop nonequilibrium  perturbation theory for inelastic electron current through a quantum dot.
We derive  Lindblad type kinetic equation for an embedded quantum dot (i.e. a quantum dot connected to Lindblad dissipators through a buffer zone). The kinetic equation is converted to non-Hermitian field theory in Liouville-Fock space.
The general nonequilibrium many-body perturbation  theory is developed and applied to the quantum dot with electron-vibronic and electron-electron interactions.
Our perturbation theory becomes equivalent to Keldysh nonequilibrium Green's functions perturbative treatment provided that the buffer zone is large enough to alleviate the problems associated with approximations of the Lindblad kinetic equation.
\end{abstract}

\section{Introduction}\label{introduction}

Study of the electron transport through nanoscopic systems remains one of the most active areas of contemporary condensed matter physics.
Most of the theoretical research has been done so far with the use of Keldysh nonequilibrium Green's functions (NEGF)~\cite{keldysh65} and scattering theory based approaches~\cite{imry1999}. NEGF applications to electron transport were pioneered by Caroli et al.\cite{Caroli71} in early 1970s.
Keldysh NEGF become particularly useful in the development of systematic perturbation theories for electron-vibronic and electron-electron interactions in the current-carrying
nanosystem.
In particular, nonequilibrium effects originated from electron-vibration coupling have attracted a lot of attention recently because of their importance in single-molecule
electronics~ \cite{galperin:035301,thoss11,PhysRevB.69.245302,dahnovsky:014104,Dash2010}.
Various kinds of perturbation theories to deal with electronic correlations have been also recently developed~\cite{PhysRevB.80.165305,schmitt,PhysRevB.75.075102,PhysRevB.77.115333,PhysRevB.79.155110,thygesen07}.

The electron transport through the system of interacting electrons (either with themselves or with some vibrational fields)   involves two different
energy scales:  One energy scale is related to  the tunneling coupling between the nanosystem and macroscopic leads and the second one is the strength of  the interactions inside the nanosystems.
NEGF usually treats the tunneling interaction exactly, but it has to rely on various types of perturbative calculations to account for correlations. On the other hand, the approaches based on kinetic equations are able to  treat the correlations inside the nanosystem very accurately (even exactly in the case of simple model systems) but the tunneling part is usually taken into account in the second or sometimes higher orders perturbation theory~\cite{gurvitz96,PhysRevB.78.235424,PhysRevB.74.235309,PhysRevB.80.045309,PhysRevB.71.205304,PhysRevB.72.195330,ovchinnikov:024707}. This immediately rules out the application of kinetic equations to the one of the most interesting transport  regimes  when there is no energy scale separation between coupling to the electrode and the correlations in the the systems (in other words,
to the case  when the tunneling time for electron  becomes comparable with the characteristic time for the development of  correlations in the dot).

 Our approach to the use of kinetic equations for electron transport is different and will be elaborated in details in the Sec.~\ref{Lindblad_kinetic_equation}.  We begin with relatively simple kinetic equation of the Lindblad type, but we make it exact for the nonequilibrium steady state by the introduction of  the finite buffer zones between the quantum dot and macroscopic leads
 (so called embedding of the quantum dot)~\cite{dzhioev11a,dzhioev11b,stability}.
 To fully link transport kinetic equations with the many-body methods we transform it to Liouville-Fock (or super-Fock) space and it becomes equivalent to effective non-Hermitian field theory with the right vacuum vector, which corresponds to nonequilibrium steady state density matrix.  This combination of the embedding and the use of Liouville-Fock space enables us to overcome the usual  limitations of the kinetic equation based approaches.  The main goal of the paper is mostly methodological. Namely, we develop nonequilibrium perturbation theory in terms of electron-vibronic and electron-electron interaction and test our theory against the NEGF results obtained for out of equilibrium local Holstein and Anderson models.

The rest of the paper is organized as follows.
In Sec.~\ref{Lindblad_kinetic_equation}, we derive the Lindblad equation for embedded quantum dot and discuss the underlying approximations.
In Sec.~\ref{Lindblad_kinetic_equation}, we also  describe superoperator formalism and convert the kinetic equation to non-Hermitian field theory in Liouville-Fock space.
Section~\ref{Perturbative_calculations} presents
the main equations of nonequilibrium  many-body perturbation theory, applications to local Holstein and Anderson models, and comparison with NEGF.
Conclusions are given in Sec.~\ref{conclusion}. We use natural units throughout the paper: $\hbar= k_B = |e| = 1$, where $-|e|$ is the electron charge.

\section{Lindblad kinetic equation for embedded quantum system in Liouville-Fock space}\label{Lindblad_kinetic_equation}

\subsection{Lindblad kinetic equation for embedded quantum dot}
\begin{figure}[t!]
\begin{center}
\includegraphics[width=1.\columnwidth]{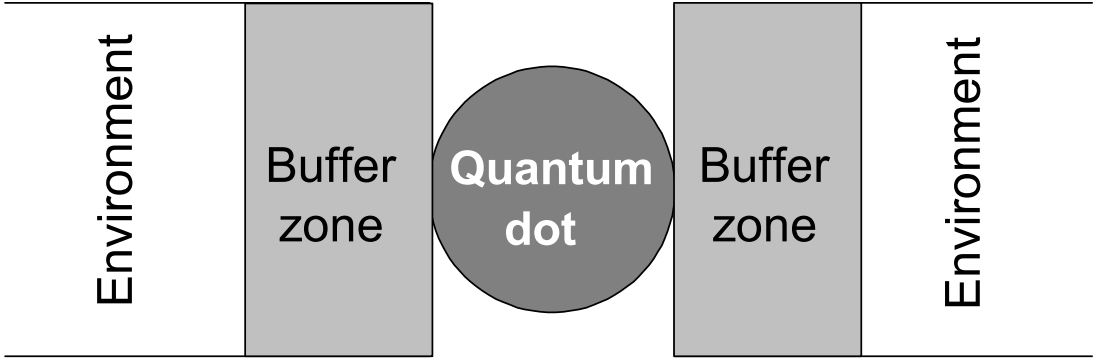}
\end{center}
\caption{Schematic illustration of quantum dot embedding. The electrodes are divided into macroscopic "environment"  and buffer zone. The projection of the environment  results into the Lindblad kinetic equation for the reduced density matrix of the buffer and quantum dot. Each buffer zone
contains a finite number of discrete single-particle levels }
\label{system}
\end{figure}

We begin by considering a quantum system (e.g. quantum dot, molecule, etc)  connected to two electrodes, left and right, with different chemical potentials.
Each electrode is partitioned into two parts (Fig~\ref{system}): the macroscopically large lead (environment)  and the finite buffer zone between the
system and the environment. So the Hamiltonian of the whole system is
\begin{equation}
{\cal H} = H_{S}+ H_{SB} + H_B + H_{BE} + H_E .
\label{h}
 \end{equation}
We assume that the environment and the buffer zones are described by the noninteracting Hamiltonians
\begin{equation}
{  H}_E = \sum_{  k \alpha  }  \varepsilon_{k\alpha } a^{\dagger}_{ k \alpha  } a_{ k \alpha },~~
{  H}_B = \sum_{ b \alpha}  \varepsilon_{ b\alpha } a^{\dagger}_{b \alpha  }  a_{b \alpha }.
\end{equation}
Here $\varepsilon_{k\alpha}$ denote the continuum single-particle spectra of the  left ($\alpha = L$) and  right ($\alpha = R$)   lead states,
 $a^{\dagger}_{k \alpha }$ ($a_{k \alpha  }$) create (annihilate) electron in the lead state
$ {k \alpha }$.
The buffer zones have discrete energy spectrum $\varepsilon_{b\alpha }$ with corresponding creation and annihilation operators
$a^\dag_{b \alpha }$ and $a_{b \alpha}$.
The  system Hamiltonian is taken in the most general form:
\begin{equation}\label{H_S}
  H_S=\sum_{s   } \varepsilon_s a^\dag_{s  } a_{s  } + H_S^{\prime},
\end{equation}
where  $a^{\dagger}_{s }$ ($a_{s  }$) create (annihilate) electron in the single-particle state
$ \varepsilon_s$ in the dot and $H_S^{\prime}$  contains two-particle electron-electron correlations, and/or
electron-vibration coupling. The buffer-environment and quantum dot-buffer couplings  have the standard tunneling form:
\begin{equation}
{  H}_{BE} = \sum_{ b k \alpha}  ( v_{b k \alpha } a^{\dagger}_{b \alpha   }  a_{k \alpha  } + \mathrm{h.c.}) ,
\label{v}
\end{equation}
\begin{equation}
{H}_{SB} = \sum_{ s b \alpha  }  ( t_{ s b \alpha } a^{\dagger}_{b \alpha } a_{s  } + \mathrm{h.c.} ).
\label{t}
\end{equation}

Now we introduce an embedded system which consist of the quantum system itself and the buffer zones.  We have recently demonstrated that if we take the buffer zones
sufficiently large the density matrix of the embedded system obeys the kinetic equation of Lindblad type.
The technical details of the derivations and underlying approximations can be found in Appendix of~\cite{stability}. Here we give only the sketch of the derivation with the emphasis on important physics relevant to our subsequent discussion.

The starting point is the Liouville equation for the total density matrix $\chi(t)$ in the interaction
picture
\begin{equation}\label{eq__tdm1}
\dot{{\chi}}_I(t) = -i[v_I(t), \chi_I(t)].
\end{equation}
Here the  buffer-environment coupling $H_{BE}$ is treated as an interaction Hamiltonian, i.e., ${\mathcal H} = h+H_{BE}$ and $v_I(t)=e^{iht}H_{BE}e^{-iht}$.
To derive the Lindblad master for the reduced density matrix of the embedded system, $\rho_I(t)=\mathrm{Tr}_E \chi_I(t)$,
we take the trace over  the environment in Eq.~\eqref{eq__tdm1} and make the following approximations:
\begin{enumerate}
  \item The total density matrix can be factorized as  $\chi_I(t) = \rho_I(t)\rho_E$ , where $\rho_E$ is density matrix of the environment taken in the equilibrium grand canonical ensemble form (Born approximation);

  \item The environment relaxation time is very fast,  so we can use local-time (Markov) approximation for the reduced density matrix;

  \item The single particle states in the buffer zone propagate as  free states
  \begin{equation}
    e^{iht}a_{b\alpha} e^{-iht} = e^{-i\varepsilon_{b\alpha}t} a_{b\alpha} +O(1/\sqrt{N})
    \label{sqrtn}
  \end{equation}
   where $N$ is the number of discrete single particle levels of the buffer zone;

  \item Rapidly oscillating terms proportional to $\mathrm{exp}[i(\varepsilon_{b\alpha}-\varepsilon_{b'\alpha})]$ for $\varepsilon_{b\alpha}\ne\varepsilon_{b'\alpha}$ are neglected (rotating wave approximation).
\end{enumerate}

Under these approximations, the Liouville equation~\eqref{eq__tdm1} reduces to a master equation for the reduced density matrix in Lindblad form.
In the Schr\"{o}dinger representation it can be written as
\begin{align}
\label{lindblad}
&  \frac{d\rho(t)}{dt} =-i[H,\rho(t)] + {\Pi}\rho(t).
\end{align}
Here the Hamiltonian $H$ includes the Lamb shift of the single-particle levels of  the buffer zones
\begin{equation}
H= H_S +H_{SB} +H_B +\sum_{b \alpha }\Delta_{b\alpha}   a^{\dagger}_{b \alpha} a_{b \alpha},
\end{equation}
and the non-Hermitian dissipator is given by the standard Lindblad form
\begin{align}\label{non_herm}
{\Pi}\rho(t) = \sum_{b\alpha} \sum_{\mu=1,2} \bigl(2 L_{ b\alpha   \mu}\rho(t) L^\dag_{ b\alpha  \mu} - \{L^\dag_{b \alpha \mu} L_{ b \alpha  \mu},\rho(t) \}\bigr).
\end{align}
The operators $L_{b\alpha 1}$ and $L_{b\alpha 2}$ are referred to as the Lindblad  operators, which represent the buffer-environment interaction. They have the following
form:
\begin{align}
  L_{b\alpha 1} = \sqrt{\Gamma_{b\alpha 1}}a_{b \alpha},~~L_{b \alpha  2} = \sqrt{\Gamma_{b\alpha 2}  }a^\dag_{ b  \alpha}.
\end{align}
with $\Gamma_{b\alpha 1}=\gamma_{b\alpha}(1-f_{b\alpha})$,  $\Gamma_{b\alpha 2}=\gamma_{b\alpha}f_{b\alpha}$. Here $ f_{b\alpha} =[1+ e^{\beta(\varepsilon_{b\alpha} - \mu_\alpha) }]^{-1}$
and   $\gamma_{b\alpha} $ ($\Delta_{b\alpha}$) is the imaginary (real)  part of the environment self energy
$\sum_k |v_{bk\alpha }|^2/(\varepsilon_{b\alpha}-\varepsilon_{k\alpha } +i0^+)$.

The Lindblad master equation describes the time evolution of the open embedded quantum system preserving the probability and the positivity of the density
matrix. Open boundary conditions are taken into account by the non-Hermitian dissipative part of Eq.\eqref{lindblad}, $\Pi \rho(t)$, which represents the
influence of environment on the buffer zone.  The applied bias potential enters into Eq.\eqref{lindblad}
via fermionic occupation numbers $f_{b\alpha}$ which depend on the temperature ($\beta=1/T$) and the chemical potential $\mu_\alpha$ in the left
and right electrodes.

\subsection{Liouville-Fock space}

Let us convert the  Lindblad master equation~\eqref{lindblad} to a non-Hermitian field theory suitable for perturbative many-body calculations. To this aim we need to introduce the concept of creation and annihilation superoperators
acting on the Liouville-Fock space~\cite{schmutz78,prosen08,Harbola2008,dzhioev11a}. Our introduction of the Liouville-Fock space closely follows Schmutz work~\cite{schmutz78}. It is general and not restricted to the particular choice of the kinetic equation.


Let $\{ |n)\}$ be a complete orthonormal basis set in the Fock space  ${\cal F}$
\begin{equation}\label{compl}
  \sum_n |n)  (n| = I,\;  \;\;  (n| m ) = \delta_{nm}.
\end{equation}
It is formed by particle number  eigenstates $|n )= a^\dag_{j_1}\ldots a^\dag_{j_n} |0)$, such that $a^\dag_j a_j |n) = n_j |n)$. Here
$|0)$ is the vacuum state and $a_j^\dag$, $a_j$  are creation and annihilation operators for single-particle state $j$. Without loss of generality we focus on fermions, so we assume that $a_j^\dag$ and  $a_j$  satisfy the canonical anti-commutation relations.

The set of linear operators $\{A(a^\dag, a)\}$ acting on $\cal F$   form  a linear vector
space, which is called the Liouville-Fock space associated with $\cal F$. We denote an element of the  Liouville-Fock space by $\ket{A}$.
The scalar product of two elements of the Liouville-Fock space is defined as
\begin{equation}
  \braket{A_1|A_2} = \mathrm{Tr}(A_1^\dag A_2).
\end{equation}
In the Liouville-Fock space we introduce a complete orthonormal basis $ \{\ket{m,n} = \ket{|m)(n|}\}$, which satisfies
\begin{equation}
 \bra{mn}m'n'\rangle= \delta_{mm'}\delta_{nn'}, ~~ \sum_{mn} \ket{mn}\bra{mn}=\bar I.
\end{equation}
Here $\bra{mn} = \ket{mn}^\dag = \bra{[|m)(n| ]^\dag}=\bra{|n)(m|}$, and $\bar I$ is the identity operator in the Liouville-Fock space.
Then, for an arbitrary element of the Liouville-Fock space we have
\begin{equation}
\ket{A} = \sum_{mn} A_{mn} \ket{mn},
\end{equation}
where $A_{mn}=\braket{m|A|n}=\braket{mn|A}$. In particular, the  identity operator $I$ in Eq.~\eqref{compl} corresponds to
\begin{equation}\label{unitoperator}
  \ket{I} = \sum_n\ket{n,n}.
\end{equation}
The scalar product of a vector $\ket{A}$ with $\bra{I}$   is equivalent to the trace operation in the Fock space,
\begin{equation}\label{tr}
\braket{I|A}=\mathrm{Tr}(A),
\end{equation}
and for the density matrix we have $\braket{I|\rho}=1$.

As was suggested by Schmutz~\cite{schmutz78} we introduce superoperators $\hat a$, $\widetilde a$
through their action on the basis vectors $\ket{mn}$
\begin{equation}\label{an_sup}
  \hat a_j \ket{mn} = \ket{ a_j  | m)(n|},~~\widetilde a_j \ket{mn} = i(-1)^{\mu} \ket{ | m)(n|a^\dag_j},
\end{equation}
where $\mu = \sum_j(m_j + n_j)=m+n$. By analyzing the Hermitian conjugate of the  matrix elements of $\hat a$, $\widetilde a$, we find
\begin{equation}\label{cr_sup}
  \hat a^\dag_j \ket{mn} = \ket{ a^\dag_j | m)(n| },~~\widetilde a^\dag_j \ket{mn} =  i(-1)^\mu \ket{ | m)(n|a_j}.
\end{equation}
It follows from~\eqref{an_sup} and~\eqref{cr_sup}  that superoperators  $\hat a$, $\hat a^\dag$ simulate the action of $a$ and $a^\dag$ on $|m)(n|$ from the left, while
$\widetilde a$, $\widetilde a^\dag$ simulate the action of $a^\dag$ and $ a$ on $|m)(n|$ from the right.
Here we would like to emphasize that our definition of tilde superoperators $\widetilde a$, $\widetilde a^\dag$
differs from Schmutz's definition by phase factors $-i$ and $+i$, respectively.
The reason for introducing these factors is that the so-called tilde-substitution rule (see bellow) becomes simpler.
We also note that the alternative definition for  superoperators is used in~\cite{Harbola2008}, where the  "right" creation and annihilation superoperators are not Hermitian conjugate to each other.

As follows from \eqref{an_sup} and \eqref{cr_sup}, the superoperators $\hat a_j$, $\widetilde a_j$ , $\hat a^\dag_j$, $\widetilde a^\dag_j$
obey the fermionic anti-commutation relations:
\begin{equation}
  \{\hat a_i, \hat a^\dag_j\}=\{\widetilde a_i, \widetilde a^\dag_j\} = \delta_{ij},
  \end{equation}
while other anti-commutators vanish
\begin{equation}
  \{\hat a_i, \hat a_j\}=\{\widetilde a_i, \widetilde a_j\} = \{\hat a_i, \widetilde a_j\} = \{\hat a_i, \widetilde a^\dag_j\} = 0.
  \end{equation}
It also follows from \eqref{an_sup} and \eqref{cr_sup} that
$\hat a\ket{00} = \widetilde a\ket{00}=0$ and the Liouville-Fock space basis vectors are generated from the vacuum $\ket{00}$ by
application of the creation superoperators
\begin{equation}\label{LF_basis}
  \ket{mn} =  (-i)^{n^2} \hat a^\dag_{k_1}\ldots \hat a^\dag_{k_m}\widetilde a^\dag_{l_1}\ldots \widetilde a^\dag_{l_n}\ket{00}.
\end{equation}
Moreover, basis vectors $\ket{mn}$ are "superfermion" number eigenstates
\begin{equation}
  \hat a_j^\dag \hat a_j \ket{mn} = m_j \ket{mn},~~\widetilde a_j^\dag \widetilde a_j \ket{mn} = n_j \ket{mn}.
\end{equation}

Using the definition of superoperators we can rewrite the identity~\eqref{unitoperator} in the following form
\begin{equation}\label{vector I}
  \ket{I} = \exp(-i\sum_j \hat a^\dag_j \widetilde a^\dag_j)\ket{00}.
\end{equation}
Note, that because of the different  definition of tilde superoperators, the obtained expression for $\ket{I}$
differs from Schmutz's analogous expression~\cite{schmutz78} by the phase factor $(-i)$ in the exponent. From~(\ref{an_sup},\ref{cr_sup})  and~\eqref{vector I} we find
that  the superoperators $\hat a^\dag_j$ and $\hat a_j$ are connected to their tilde conjugate
$\widetilde a^\dag_j$ and $\widetilde a_j$  by the relations
\begin{equation}\label{rule_1}
\hat a_j  \ket{I} = -i \widetilde a^\dag_j \ket{I},~~ \hat a^\dag_j  \ket{I} = -i \widetilde a_j \ket{I}.
\end{equation}

For an operator $A=A(a^\dag, a)$ given by the power series of creation and annihilation operators  we define two superoperators
\begin{equation}\label{superop}
 \hat A= A(\hat a^\dag, \hat a),~~~\widetilde A=A^*(\widetilde a^\dag, \widetilde a).
\end{equation}
Here, the $*$ means  the complex conjugate of the $c$-number coefficients.
The relation between non-tilde and tilde superoperators is given by the following tilde conjugation rules
\begin{align}\label{TC_rules}
 (c_1 \hat A_1 + c_2 \hat A_2)\widetilde{} = c_1^* \widetilde{A}_1 + c^*_2 \widetilde {A}_2, ~~
 (\hat A_1\hat A_2)\widetilde{}=\widetilde{A}_1\widetilde{A}_2, ~~ (\widetilde A)\widetilde{} = A.
\end{align}
Applying tilde conjugation to $\ket{mn}$ we find
\begin{equation}
  \ket{mn}\widetilde{}=(+i)^{\mu^2}\ket{nm},
\end{equation}
where $\mu=m+n$. Therefore $\ket{I}\widetilde{}=\ket{I}$, i.e., $\ket{I}$ is tilde-invariant.
Generally, if $A=A(a^\dag, a)$ is a Hermitian bosonic operator then $\ket{A}\widetilde{}=\ket{A}$.

According to the definition of the superoperator $\hat A$, if $A=\sum_{mn}A_{mn}|m)(n|$ then  $\hat{A} =\sum_{mnk} A_{mn}\ket{mk}\bra{nk}$ and
we  obtain
\begin{equation}\label{ket_A1}
  \ket{A} = \hat A\ket{I},~~\ket{A}\widetilde{} = \widetilde A\ket{I},
\end{equation}
\begin{equation}\label{ABI_1}
\ket{A_1A_2} =  \hat A_1 \hat A_2\ket{I} =  \hat A_1 \ket{A_2}.
\end{equation}
Therefore, the expectation value of an operator $A=A(a^\dag, a)$ in the state with the density matrix
$\rho=\rho(a^\dag, a)$ can be calculated as the matrix element of the corresponding superoperator $\hat A=A(\hat a^\dag, \hat a) $
sandwiched between $\bra{I}$ and $\ket{\rho}=\hat\rho\ket{I}$
\begin{equation}
  \braket{A}=\mathrm{Tr}(A\rho)=\braket{I|A\rho} =\braket{I|\hat A|\rho}.
\end{equation}
Using~\eqref{rule_1} we can show that the following tilde-substitution rule is valid
\begin{equation}\label{TSR}
\hat A\ket{I} = \sigma_A \widetilde A^\dag \ket{I}.
\end{equation}
Here $\sigma_A = +1$ if $A$ is  a bosonic operator and $\sigma_A= -i $ if $A$ is  a fermionic operator.  Moreover, taking into account that non-tilde and tilde
fermion superoperators anti-commute  we find that
\begin{equation}\label{ABI_2}
\hat A_1 \ket{A_2}= i  \widetilde A_2^\dag \ket{A_1},
\end{equation}
if both $A_1$ and $A_2$ are fermionic operators, and
 \begin{equation}\label{ABI_3}
  \hat A_1 \ket{A_2} = \sigma_{A_2}  \widetilde A_2^\dag \ket{A_1}
\end{equation}
otherwise. It should be noted that Schmutz tilde-substitution rule~\cite{schmutz78}
is cumbersome and it takes the simple form like~\eqref{TSR} only if all terms in the power series of $A(a^\dag, a)$ have the common quantity $m-n$. Here $m (n)$ is the number of
creation (annihilation) operators.

The general prescription to obtain equation for $\ket{\rho(t)}$ from the kinetic equation for $\rho(t)$ is the following. First,
we transform the kinetic equation for $\rho=\rho(a^\dag, a)$ into the kinetic equation for  $\hat\rho = \rho(\hat a^\dag, \hat a)$ by formally replacing
all operators $a^\dag,~a$ by superoperators $\hat a^\dag,~\hat a$. Then, we multiply the kinetic equation from the right on vector $\ket{I}$  and use~\eqref{TSR}-\eqref{ABI_3} to convert the kinetic equation to
the Schr\"{o}edinger-like equation for the vector $\ket{\rho(t)}=\hat\rho(t)\ket{I}$:
\begin{equation}\label{sch}
  i \frac{d}{d t}\ket{\rho(t)} = L(\hat a^\dag, \hat a, \widetilde a^\dag, \widetilde a)\ket{\rho(t)},
\end{equation}
where $L$ is the Liouvillian which depends on both non-tilde and tilde superoperators.
Particularly,    the Liouvillian for the Lindblad master equation~\eqref{lindblad}
becomes
\begin{equation}
L = \hat H - \widetilde H-i \sum_{b\alpha}\Pi_{b\alpha},
\label{H_lind}
\end{equation}
where
\begin{align} \label{Pi}
\Pi_{b\alpha} =&(\Gamma_{b\alpha1}-\Gamma_{b\alpha2})(\hat a^\dag_{b\alpha}\hat a_{b\alpha}+
  \widetilde a^\dag_{b\alpha}\widetilde a_{b\alpha})
  \notag\\&-
  2i(\Gamma_{b\alpha1}\widetilde a_{ b\alpha}\hat a_{b \alpha}+\Gamma_{b\alpha2}\widetilde a^\dag_{ b\alpha}\hat a^\dag_{ b\alpha})+2\Gamma_{b\alpha2}.
   \end{align}
In derivation of~\eqref{H_lind} and~\eqref{Pi} we took into account that $\hat\rho = \rho(\hat a^\dag, \hat a)$ is a bosonic superoperator which commutes with all tilde superoperators.
Due to the Lindblad dissipators, the Liouville superoperator~\eqref{H_lind} is non-Hermitian. In addition, as $\ket{\rho}$ is tilde-invariant,   the Liouvillian obeys the property
$(L)\widetilde{} =  -L$.

Taking the time derivative of $\braket{I|\rho(t)}=1$ we find that $\bra{I} L=0$, i.e., $\bra{I} $ the left
zero-eigenvalue eigenstate of the Liouvillian superoperator. Since also $\bra{I}$ is the vacuum for $\hat a^\dag_j-i \widetilde a_j$ and
$\widetilde a^\dag_j + i \hat a_j$ superoperators, it is appropriate to call $\bra{I}$ left vacuum vector.
For the electron transport problem we focus on nonequilibrium steady-state
where the current through the quantum dot is given by
\begin{equation}
  \braket{J_\alpha} =\mathrm{Tr}(J_\alpha\rho_\infty) = \braket{I|\hat J_\alpha|\rho_\infty}.
\end{equation}
Here $\hat J_\alpha$ is  the current superoperator, and
the stationary, steady-state solution of~\eqref{sch}, $\ket{\rho_\infty}$, is the right zero-eigenvalue eigenstate (right vacuum vector) of the Liouville superoperator
\begin{equation}\label{lrho=0}
  L\ket{\rho_\infty}=0.
\end{equation}
In the next section, we show how one can find $\ket{\rho_\infty}$ perturbatively starting from the free-field approximation for nonequilibrium density
matrix.

\section{Perturbative calculations of the steady state density matrix and electron current}\label{Perturbative_calculations}

\subsection{Nonequilibrium many-body perturbation theory}

Let us make the important remark on the notation use in the rest of the paper: only creation/annihilation operators written with letters $a,d$ (such as for example $a_{b \alpha}$ and $a^{\dagger}_{b \alpha}$) are related to each other by the Hermitian conjugation; all other creation  $c^{\dagger}, b^{\dag}, \gamma^\dag$ and annihilation $c,b, \gamma$ operators are "canonically conjugated" to each other, i.e., for example,  $c^{\dagger}$ does not mean $(c)^{\dagger}$ although $ cc^{\dagger} \pm c^{\dagger} c =1$ ($\pm$ - bosons/fermions). We will also use the same notation for the non-tilde superoperators
$\hat a^\dag_j$ and  $\hat a_j$ as the ordinary operators $ a^\dag_j$ and  $a_j$ bearing in mind that all operators acting in the Liouville-Fock space are
are superoperators.

We start by rewriting the Liouvillian~\eqref{H_lind} as
\begin{equation}
  L=L^{(0)}+L',
\end{equation}
where $L^{(0)}$ is the quadratic unperturbed part of $L$, and
\begin{equation}
L'=H_S^{\prime}-\widetilde H_S^{\prime}
\end{equation}
 is a perturbation. Then  using the equation of motion method
\begin{align}\label{eq_mot}
[c^\dag_n,L^{(0)}]&=-\Omega_n c^\dag_n,
\notag\\
[c_n,L^{(0)}]&=\Omega_n c_n,
\end{align}
we  exactly diagonalize\cite{dzhioev11a}  $L^{(0)}$ in terms of nonequilibrium quasiparticle
creation and annihilation operators:
\begin{equation}
  L^{(0)} = \sum_{n}  (\Omega_n c^\dag_{n}c_{n} - \Omega^*_n \widetilde c^\dag_{n}\widetilde c_{n}).
\end{equation}
Here $\widetilde c^\dag_{\sigma n}, \widetilde c_{\sigma n} $ are obtained from $c^\dag_{\sigma n}, c_{\sigma n} $ by the tilde conjugation rules.

The nonequilibrium quasiparticle creation and annihilation operators are connected to $a^\dag,~a,~\widetilde a^\dag,~\widetilde a$ by
canonical (but not unitary) transformations:
\begin{align}\label{b_to_c}
 c^\dag_{n} &= \sum_s \psi_{n,s} b^\dag_s +\sum_{b\alpha}\psi_{n,b\alpha} b^\dag_{b\alpha} ,
  \notag\\
 c_{n} &=\sum_s(\psi_{n,s} b_s +  i  \varphi_{n,s} \widetilde b^\dag_s)
   +\sum_{b\alpha}(\psi_{n,b\alpha}  b_{b\alpha} +i \varphi_{n,b\alpha}
  \widetilde b^\dag_{b\alpha}),
\end{align}
where
\begin{align*}
&b^\dag_s= a^\dag_s - i \widetilde a_s,~~ b_s=a_s,~~b^\dag_{b\alpha}=a^\dag_{b\alpha} - i \widetilde a_{b\alpha},
\notag\\
&b_{b\alpha}=(1-f_{b\alpha})a_{b\alpha}+if_{b\alpha}\widetilde a^\dag_{b\alpha}.
\end{align*}
Nonequilibrium quasiparticle creation and annihilation operators
obey the fermionic anti-commutation relations. In particular, from $\{c_n, c^\dag_{n'}\}=\delta_{nn'}$ and $\{c_n, \widetilde c_{n'}\}=0$
we find the following orthonormality conditions for amplitudes
\begin{align}\label{orthonormality}
  &\sum_s \psi_{n,\,s}\psi_{{n'},\,s}+\sum_{b\alpha} \psi_{n,\,b\alpha}\psi_{{n'},\,b\alpha}=\delta_{nn'},
     \notag \\
 & \sum_s (  \psi_{n,\,s}\varphi^*_{{n'},\,s}-\varphi_{n,\,s}\psi^*_{{n'},\,s})
  \notag \\
 & ~~~~~  +\sum_{b\alpha}(  \psi_{n,\,b\alpha}\varphi^*_{{n'},\,b\alpha}-\varphi_{n,\,b\alpha}\psi^*_{{n'},\,b\alpha})=0.
\end{align}

By construction,  $\langle I|$ is the left  vacuum for $c^\dag_{n},~\widetilde c^\dag_{n}$ operators.
The vacuum  for  $c_{n},~\widetilde c_{n}$ operators, $\ket{\rho^{(0)}_\infty}$, is automatically
the  zero-eigenvalue eigenstate of the unperturbed Liouvillian $L^{(0)}$, i.e., it is
the steady state density matrix in the zeroth-order approximation:
\begin{equation}
  L^{(0)}\ket{\rho_\infty^{(0)}}=0,~~~\bra{I}\rho_\infty^{(0)}\rangle=1.
\end{equation}
In other words, the the zeroth-order  density matrix is the density matrix which does not contain nonequilibrium quasiparticle excitations.

Now we introduce the continuous real parameter $\lambda$, which
will be set to unity in the end of the calculations,
\begin{equation}
  L =  L^{(0)} + \lambda L'
\end{equation}
and expand  the exact steady state density matrix in powers of $\lambda$,
\begin{equation}\label{ss_expansion}
  \ket{\rho_\infty} =\sum_{p=0}\lambda^{p}\ket{\rho^{(p)}_\infty}.
\end{equation}
Substituting (\ref{ss_expansion}) into Eq.~\eqref{lrho=0},   we obtain
equation for the  $p$th-order correction to the zeroth-order density matrix:
\begin{equation}\label{pert_eq}
 L_{0} \ket{\rho^{(p)}_\infty} = - L'  \ket{\rho^{(p-1)}_\infty}.
\end{equation}
or $\ket{\rho^{(p)}_\infty} = (- L_{0}^{-1}L')^p\ket{\rho^{(0)}_\infty}$. Here, $L'$ is expressed in terms of the  nonequilibrium quasiparticles.
Thus, starting from $\ket{\rho^{(0)}}$ we can find any $p$th-order corrections to it. In addition,
${\bra{I}\rho_\infty^{(p)}\rangle=0}$ for $p\ge1$ since $\ket{\rho^{(p)}_\infty}$ contains excited nonequilibrium
quasiparticles.

To calculate the current through the quantum dot we express the current superoperator
\begin{equation}
 J_\alpha = -i \sum_{bs}t_{sb\alpha}( a^\dag_{b\alpha} a_s -  a^\dag_s a_{b\alpha})
 \end{equation}
in terms of nonequilibrium quasiparticle creation and annihilation operators and compute its expectation
value with respect to $\bra{I}$ and $\ket{\rho_\infty}$. As a result we get the following expansion
\begin{align}\label{current}
J_\alpha = \sum_{p=0} \lambda^{p}J^{(p)}_\alpha.
\end{align}
Here, $J^{(0)}_\alpha$ is zeroth-order current for the system without interaction
\begin{align}\label{current_HF_And1}
  J^{(0)}_\alpha = -2\mathrm{Im}\sum_{bs n}t_{sb\alpha}\psi_{n,b\alpha}\varphi_{n,s},
\end{align}
and   $J_\alpha^{(p)}$ is the $p$th-order correction to it
\begin{align}\label{current_p}
  J_\alpha^{(p)}
  = -2\mathrm{Im}\sum_{b smn}t_{sb\alpha}\psi^*_{n,b\alpha}\psi_{m,s}F^{(p)}_{mn},
\end{align}
where $F^{(p)}_{mn}$ is the expansion coefficient in
\begin{equation}
  \ket{\rho^{(p)}_\infty}=i\sum_{mn} F^{(p)}_{mn} c^\dag_{m}\widetilde c^\dag_{n}\ket{\rho^{(0)}_\infty}+\ldots
\end{equation}
and $F^{(p)}_{mn}=(F^{(p)}_{nm})^*$ as follows from  $\ket{\rho^{(p)}_\infty}={ \ket{\rho^{(p)}_\infty}}\widetilde{}$~.
Thus, the problem of computing the $p$th-order correction to the unperturbed current is reduced to finding $F^{(p)}_{mn}$ by solving
 Eq.~\eqref{pert_eq}.

Using the same method we can obtain perturbative expansion for  the population of a quantum dot single-particle level
\begin{equation}
  n_s = \bra{I} a^\dag_s a_s\ket{\rho_\infty} = \sum_{p=0} n_s^{(p)},
\end{equation}
where
\begin{equation}
 n_s^{(0)}  = \sum_n \psi_{n,s} \varphi_{n,s}, 
 ~~~
  n_s^{(p)} = - \sum_{mn}\psi_{m,s}\psi^*_{n,s}F^{(p)}_{mn}.
\end{equation}
The anti-commutation condition $\{b_s,\widetilde b_s\}=0$ imposes the constraint on the amplitudes from which  follows that  $n_s^{(0)}$ is a real number.

\subsection{Electron-vibronic coupling}\label{EVC}

As the first application of the  method we consider the Hamiltonian $H_S$ which describes one electronic single-particle level coupled linearly to
a  vibration mode (phonon)  of frequency $\omega_0$ (so-called local Holstein model)
\begin{equation}\label{H_s_vibr}
  H_S=\varepsilon_0 a^\dag a + \omega_0 d^\dag d + \kappa a^\dag a (d^\dag + d).
\end{equation}
For simplicity we assume that the tunneling matrix element in Eq.~\eqref{t} is real number $t$ independent of indices $\alpha$ and $b$. The electron spin does not play any role here, so we suppress the spin index in the equations in this section.
Replacing $\kappa$ by $\lambda\kappa$, we arrive to
perturbation expansion of the steady state density matrix $\ket{\rho_{\infty}}$ with respect to electron-vibronic coupling
\begin{equation}
  \ket{\rho_{\infty}} = \sum_{p=0}\lambda^p \ket{\rho_{\infty}^{(p)}}.
\end{equation}

To find the zeroth-order density matrix $\ket{\rho_{\infty}^{(0)}}$, we diagonalize the fermionic part of $L^{(0)}$.
The resulting creation and annihilation operators have the form~\eqref{b_to_c},  and   amplitudes $\psi,~\varphi$ satisfy the
following system of equations
\begin{equation}\label{sys1}
\left\{
\begin{array}{l}
 \varepsilon_0 \psi_{n}-t \sum\limits_{b\alpha}\psi_{n,b\alpha}=\Omega_n \psi_{n}
  \\
 E_{b\alpha}\psi_{n,b\alpha}-  t\psi_{n}=\Omega_n \psi_{n,b\alpha},
  \end{array}\right.
\end{equation}

\begin{equation}\label{sys2}
\left\{\begin{array}{l}
   (\varepsilon_0 - \Omega_n) \varphi_{n}-t \sum\limits_{b\alpha} \varphi_{n,b\alpha}=t \sum\limits_{b\alpha} f_{b\alpha} \psi_{n,b\alpha}
   \\
  (E^*_{b\alpha}-\Omega_n)\varphi_{n,b\alpha}-  t\varphi_{n}=-t f_{b\alpha} \psi_{n},
   \end{array}\right.
\end{equation}
where $E_{b\alpha}=\varepsilon_{b\alpha}-i\gamma_{b\alpha}$. The solution of eigenvalue problem~\eqref{sys1} yields
the spectrum of nonequilibrium quasiparticles, $\Omega_n$ and $-\Omega^*_n$, as well as $\psi$ amplitudes which should be normalized according to Eq.~\eqref{orthonormality}.
To find $\varphi$ amplitudes we must solve linear equations~\eqref{sys2}.

Furthermore, let $N_\omega$ be the number of vibrational quanta with frequency $\omega_0$ at temperature $1/\beta$, i.e.,
$N_\omega = ( \exp(\beta\omega_0)  - 1 )^{-1}$. When $\kappa =0$
the density matrix is factorized as $\ket{\rho_{\infty}^{(0)}} = \ket{\rho_{\infty}^{(0)}}_f \ket{\rho_{\infty}^{(0)}}_b$,
\begin{equation}
 \bra{I} d^\dag d\ket{\rho_{\infty}^{(0)}} = N_\omega.
\end{equation}
It is convenient to introduce new phonon operators
\begin{align}\label{d_to_gamma}
 &\gamma =(1+N_\omega)d - N_\omega \widetilde d^\dag,
 \notag\\
 &\gamma^\dag = d^\dag-\widetilde d
 \end{align}
 and their tilde conjugated partners, such that $\bra{I}\gamma^\dag=\bra{I}\widetilde\gamma^\dag=0$ and $\gamma\ket{\rho_{\infty}^{(0)}}=\widetilde\gamma\ket{\rho_{\infty}^{(0)}}=0$.
Then the quadratic part of the Liouvillian is diagonal in terms of introduced operators
\begin{equation}\label{L0}
  L^{(0)}= \sum_n(\Omega_n c^\dag_n c_n-\Omega^*_n \widetilde c^\dag_n \widetilde c_n) + \omega_0(\gamma^\dag\gamma - \widetilde\gamma^\dag\widetilde\gamma),
\end{equation}
and the vacuum for $c_n,~\widetilde c_n,~\gamma$, and $\widetilde\gamma$ operators is the
the zeroth-order approximation for the density matrix, $\ket{\rho_{\infty}^{(0)}}$. For the unperturbed current we have
\begin{equation}
  J^{(0)}_\alpha=-2t\mathrm{Im}\sum_{bn}\psi_{n,b\alpha}\varphi_n,
\end{equation}
while $p$th-order correction is
\begin{align}\label{current_p_Holst}
  J_\alpha^{(p)}
  = -2t\mathrm{Im}\sum_{bmn}\psi^*_{n,b\alpha}\psi_{m}F^{(p)}_{mn}.
\end{align}

To find $F^{(p)}_{mn}$  we rewrite the perturbative part of Liouvillian in terms of operators $c_n,~\gamma$, etc.:
 \begin{align}\label{Lpr}
  L'=&\sum_{mn}\Bigl\{\bigl[L^{(1)}_{mn}\gamma^\dag+L^{(2)}_{mn}\widetilde \gamma^\dag +L^{(3)}_{mn}(\gamma+\widetilde\gamma)\bigr]c^\dag_m c_n -\mathrm{t.c.}\Bigr\}
  \notag\\
  -&i\sum_{mn}\bigl[L^{(4)}_{mn}\gamma^\dag - (L^{(4)}_{nm})^*\widetilde\gamma^\dag+ L^{(5)}_{mn}(\gamma+\widetilde\gamma)\bigr]c^\dag_m\widetilde c^\dag_n
 \notag\\
  -&i\sum_{mn}L^{(6)}_{mn}(\gamma^\dag-\widetilde\gamma^\dag)c_m\widetilde c_n+\kappa n^{(0)}(\gamma^\dag-\widetilde\gamma^\dag),
\end{align}
 where coefficients $L^{(i)}$ are
\begin{align}
  L^{(1)}_{mn}&=\kappa\bigl[(\psi_m-\varphi_m)+N_\omega\psi_m\bigr]\psi_n,
  \notag\\
  L^{(2)}_{mn}&=\kappa\bigl[\varphi_m+N_\omega\psi_m\bigr]\psi_n,~~L^{(3)}_{mn}=\kappa\psi_m\psi_n,
   \notag\\
  L^{(4)}_{mn}&=\kappa\bigl[(\psi_m-\varphi_m)\varphi_n^*+N_\omega(\psi_m\varphi^*_n-\varphi_m\psi_n^*)\bigr]
   \notag\\
  L^{(5)}_{mn}&=\kappa\bigl[\psi_m\varphi_n^*-\varphi_m\psi^*_n\bigr],~~L^{(6)}_{mn}=\psi_m\psi_n^*,
\end{align}
and
\begin{equation}
n^{(0)}=\bra{I}a^\dag a\ket{\rho^{(0)}_{\infty}}=\sum_n\psi_n\varphi_n
\end{equation}
is an unperturbed electron level population. The notation 't.c.' in equation~\eqref{Lpr} means the tilde conjugation
(i.e., $c^\dag_m\to \tilde c^\dag_m$, $\gamma^\dag \to \tilde\gamma$, $L^{(1)}_{mn}\to (L^{(1)}_{mn})^*$, etc.).
Then, substituting  Eqs.~(\ref{L0}, \ref{Lpr}) into \eqref{pert_eq} we obtain the following general expression for $F^{(p)}_{mn}$
\begin{align}\label{Fp}
 F^{(p)}_{mn} = &- \frac{1}{\Omega_m-\Omega^*_n}\Bigl\{\sum_{k} L^{(3)}_{mk}\bigl[Z^{(p-1)}_{kn} + (Z^{(p-1)}_{nk})^*\bigr]
  \notag \\
  -&\sum_{k} (L^{(3)}_{nk})^*\bigl[(Z^{(p-1)}_{km})^* + Z^{(p-1)}_{mk} \bigr]-2L^{(5)}_{mn}W^{(p-1)}  \Bigr\},
\end{align}
where $Z^{(p)}_{mn}$ and $W^{(p)}$ are  coefficients in the expansion
\begin{align}\label{vibr_first}
&  \ket{\rho_{\infty}^{(p)}}=\Bigl\{W^{(p)}(\gamma^\dag+\widetilde\gamma^\dag)
   \notag\\
&  +i\sum_{mn}\bigl[Z^{(p)}_{mn}\gamma^\dag + (Z_{nm}^{(p)})^*\widetilde\gamma^\dag\bigr]c^\dag_{m}\widetilde c^\dag_n + \ldots\Bigr\}\ket{\rho_{\infty}^{(0)}}.
\end{align}

Thus, to find $p$th-order correction to the current we need first compute $Z^{(p-1)}_{mn}$ and $W^{(p-1)}$. This can be down using the same method as used to find $F^{(p)}_{mn}$.
As a result, $Z^{(p)}_{mn}$ and $W^{(p)}$ are nonvanishing only for odd $p$. Therefore, only even powers of $p$ contribute to the current expansion
as it should be for the considered model. It is interesting to note that the term $W^{(p)}(\gamma^\dag+\widetilde\gamma^\dag) \ket{\rho_{\infty}^{(0)}}$  is associated to the momentum transfer from the  electronic current to the quantum dot vibrational mode (current induced translational motion of the dot)  whereas
$Z^{(p)}_{mn}\gamma^\dag c^\dag_{m}\widetilde c^\dag_n \ket{\rho_{\infty}^{(0)}}$ and
 $ (Z_{nm}^{(p)})^*\widetilde\gamma^\dag c^\dag_{m}\widetilde c^\dag_n \ket{\rho_{\infty}^{(0)}}$ correspond to the current induced heating and cooling processes respectively.

As an example we give here explicit expressions for the first order perturbation theory $W^{(1)}$ and $Z^{(1)}_{mn}$:
\begin{equation}\label{ZW}
W^{(1)}=-\frac{n^{(0)}}{\omega_0},~~~ Z^{(1)}_{mn}=\frac{L^{(4)}_{mn}}{\Omega_m-\Omega^*_n+\omega_0}.
\end{equation}
Combining Eqs.~\eqref{ZW} and~\eqref{Fp}, we find $F^{(2)}_{mn}$. Then inserting  $F^{(2)}_{mn}$ into~\eqref{current_p_Holst} we
derive  the second-order perturbation theory correction to $J^{(0)}_\alpha$. This correction consists of two parts: the first part is proportional to $n^{(0)}$, so it
is the Hartree term, while the remaining part is the Fock term. In section~\ref{comparison}, we also verify these definitions by comparing Hartree and Fock terms
obtained within the presented approach and the exact ones given by NEGF formalism.

\subsection{Electronic correlations}

As a next example we consider electron transport through  one spin-degenerate level with local Coulomb interaction
\begin{equation}
  H_S=\varepsilon_0\sum_{\sigma}n_\sigma + Un_\uparrow n_\downarrow.
\end{equation}
Here $n_\sigma=a^\dag_\sigma a_\sigma$ is the number operator  for electrons  with spin $\sigma$ in the quantum dot.  In what follows, we again assume the tunneling matrix element is independent of $\alpha$, $b$ as well as spin $\sigma$, i.e.,
\begin{equation}
  H_{SB}=-t\sum_{\sigma b\alpha}(a^\dag_{\sigma b\alpha}a_\sigma +\mathrm{h.c}).
\end{equation}
We also assume that energy levels in the leads are spin-degenerate.

Since the quadratic part of the corresponding Liouvillian describes electron transport through
noninteractiong spin-up  and spin-down levels, it is  diagonalized by the same method as in the previous example. As a result
we obtain
\begin{equation}\label{L0_coul}
  L^{(0)}=\sum_{\sigma n}(\Omega_n c^\dag_{\sigma n} c_{\sigma n} - \Omega^*_n\widetilde c^\dag_{\sigma n} \widetilde c_{\sigma n}).
\end{equation}
The vacuum of $c_{\sigma n}$ and $\widetilde c_{\sigma n}$ operators, $\ket{\rho^{(0)}_{\infty}}$, is the density matrix in the zeroth-order perturbation theory and
\begin{equation}\label{coulomb_cur_0}
  J^{(0)}_\alpha=-4t\mathrm{Im}\sum_{bn}\psi_{n,b\alpha}\varphi_n
\end{equation}
is the corresponding current.

To find $p$th-order correction to~\eqref{coulomb_cur_0},
\begin{equation}\label{coulomb_cur_p}
J^{(p)}_\alpha=-4t\mathrm{Im}\sum_{b mn}\psi^*_{n,b\alpha}\psi_{m}F^{(p)}_{mn},
\end{equation}
we rewrite $L'=U(n_\uparrow n_\downarrow -\widetilde n_\uparrow \widetilde n_\downarrow)$ in terms of nonequilibrium quasiparticles:
\begin{align}\label{L'_coul}
L'=&\sum\limits_{\sigma kl}\Bigl\{K^{(1)}_{kl}(c^\dag_{\sigma k}c_{\sigma l}-\mathrm{t.c.}) +iK^{(2)}_{kl}c^\dag_{\sigma k}\widetilde c^\dag_{\sigma l}  \Bigr\}
&\nonumber \\
+&\sum\limits_{klmn}\Bigl\{ (L^{(1)}_{klmn}c^\dag_{k_\uparrow} c^\dag_{l_\downarrow} c_{m_\downarrow} c_{n_\uparrow} - \mathrm{t.c.})
&\nonumber \\
&+ L^{(2)}_{k l m n} c^\dag_{k\uparrow} c^\dag_{l_\downarrow} \widetilde c^\dag_{m_\uparrow} \widetilde c^\dag_{n_\downarrow}
 \nonumber \\
&+ L^{(3)}_{klmn}(c^\dag_{k_\uparrow}\widetilde c^\dag_{l_\downarrow} \widetilde c_{m_\downarrow}  c_{n_\uparrow}+
                  c^\dag_{k_\downarrow}\widetilde c^\dag_{l\uparrow} \widetilde c_{m_\uparrow}  c_{n_\downarrow}
&\nonumber \\
&~~~-c^\dag_{k_\uparrow}\widetilde c^\dag_{l_\uparrow} \widetilde c_{m_\downarrow}  c_{n_\downarrow}-
  c^\dag_{k_\downarrow}\widetilde c^\dag_{l_\downarrow} \widetilde c_{m_\uparrow}  c_{n_\uparrow})
\nonumber  \\
&+i\bigl[L^{(4)}_{klmn}(c^\dag_{k_\uparrow} c^\dag_{l_\downarrow}\widetilde c^\dag_{m_\downarrow} c_{n_\uparrow}+
  c^\dag_{k_\downarrow} c^\dag_{l_\uparrow}\widetilde c^\dag_{m_\uparrow} c_{n_\downarrow}) + \mathrm{t.c.}\bigr]
 \nonumber \\
&+i\bigl[L^{(5)}_{klmn}(c^\dag_{k_\uparrow} \widetilde c_{l_\downarrow} c_{m_\downarrow} c_{n_\uparrow}+
  c^\dag_{k_\downarrow}\widetilde c_{l_\uparrow} c_{m_\uparrow} c_{n_\downarrow})+\mathrm{t.c.}\bigr]\Bigr\}.
  \end{align}
Here $K^{(1)}_{kl}$ and  $K^{(2)}_{kl}$ are given by
\begin{align}
&K^{(1)}_{kl}=Un^{(0)}_\sigma\psi_k\psi_l,~~K^{(2)}_{kl}=-Un^{(0)}_\sigma(\psi_k\varphi_l^*-\varphi_k\psi_l^*),
\nonumber  \\
&n^{(0)}_\sigma=\bra{I} a^\dag_\sigma a_\sigma\ket{\rho^{(0)}_{\infty}}=\sum_n\psi_n\varphi_n,
\end{align}
while the coefficients  $L^{(i)}_{klmn}$ are listed in~\cite{dzhioev11b}. 

Now, substituting Eqs.~(\ref{L0_coul},\ref{L'_coul}) into~Eq.~\eqref{pert_eq}  we find the following general expression for $F^{(p)}_{mn}$
\begin{align}\label{F_p}
 &F^{(p)}_{mn}=-\frac{1}{\Omega_m-\Omega^*_n}\Bigl\{K^{(2)}_{mn}\delta_{p1}
 \notag\\
& +\sum_i \bigl[K_{mi}^{(1)}F^{(p-1)}_{in}- (K_{ni}^{(1)}F^{(p-1)}_{im})^*\bigr]
 -\sum_{ij}L^{(3)}_{mnij}F^{(p-1)}_{ji}
 \notag\\
& -\sum_{ijk}\bigl[ L^{(5)}_{mijk}G^{(p-1)}_{kjni} - (L^{(5)}_{nijk}G^{(p-1)}_{kjmi})\bigr]\Bigr\},
\end{align}
where $\delta_{p1}$ is the Kronecker delta and $G^{(p)}_{klmn}$ is a coefficient in the expansion
\begin{align}\label{G_coul}
  \ket{\rho^{(p)}_{\infty}}=\Bigl\{\sum_{klmn} G^{(1)}_{klmn}
 c^\dag_{\uparrow k}c^\dag_{\downarrow l}\widetilde  c^\dag_{\uparrow m}\widetilde c^\dag_{\downarrow n} + \ldots \Bigr\} \ket{\rho^{(0)}_\infty}.
\end{align}
In turn, the equation like \eqref{F_p} can be derived for $G^{(p)}_{klmn}$.

The exact first-order perturbation theory correction to $\ket{\rho^{(0)}_{\infty}}$ is
\begin{align}\label{rho_C_1}
  \ket{\rho^{(1)}_{\infty}}=\Bigl\{i\sum_{\sigma mn}&F^{(1)}_{mn}c^\dag_{\sigma m}\widetilde c^\dag_{\sigma n}
  \notag\\
   &+\sum_{klmn} G^{(1)}_{klmn}
 c^\dag_{\uparrow k}c^\dag_{\downarrow l}\widetilde  c^\dag_{\uparrow m}\widetilde c^\dag_{\downarrow n} \Bigr\} \ket{\rho^{(0)}_\infty},
\end{align}
 where
 \begin{align}\label{FG_1}
 F^{(1)}_{mn}=-\frac{K^{(2)}_{mn}}{\Omega_m-\Omega^*_n},~~
G^{(1)}_{klmn}= - \frac{L^{(2)}_{klmn}}{\Omega_k+\Omega_l-\Omega^*_m -\Omega^*_n}.
\end{align}
Inserting $F^{(1)}_{mn}$ into~\eqref{coulomb_cur_p} we get the first-order perturbation theory correction  $J^{(1)}_\alpha$ to the current~\eqref{coulomb_cur_0}.
This correction is proportional to $n^{(0)}$ and below we will show that it corresponds to the first-order Hartree term obtained with NEGF formalism.

Here we note, that in~\cite{dzhioev11b} we applied  perturbation theory to the  Anderson model  starting from the nonequilibrium Hartree-Fock approximation, i.e.,
$L'$ was normal ordered and did not contain quadratic terms. Therefore, in~\cite{dzhioev11b}
the mixture of two  quasiparticle configurations to $\ket{\rho^{(1)}_{\infty}}$ vanished
and the first-order perturbation theory correction to the current was zero.

To find the second-order correction to $J^{(0)}_\alpha$  we insert~\eqref{FG_1} into~\eqref{F_p}. This yields
 \begin{align}\label{F_2}
 &F^{(2)}_{mn}=-\frac{1}{\Omega_m-\Omega^*_n}\Bigl\{\sum_i \bigl[K_{mi}^{(1)}F^{(1)}_{in}- (K_{ni}^{(1)}F^{(1)}_{im})^*\bigr]
 \notag\\
 &-\sum_{ij}L^{(3)}_{mnij}F^{(1)}_{ji}-\sum_{ijk}\bigl[ L^{(5)}_{mijk}G^{(1)}_{kjni} - (L^{(5)}_{nijk}G^{(1)}_{kjmi})^*\bigr]\Bigr\}.
 \end{align}
Now, with the help of the obtained expression for $F^{(2)}_{mn}$ and Eq.~\eqref{coulomb_cur_p} we get the second-order perturbation theory correction to $J^{(0)}_\alpha$.

\subsection{Comparison with Keldysh NEGF perturbation theory}\label{comparison}

\begin{figure}[t!]
\begin{center}
\includegraphics[width=1.\columnwidth]{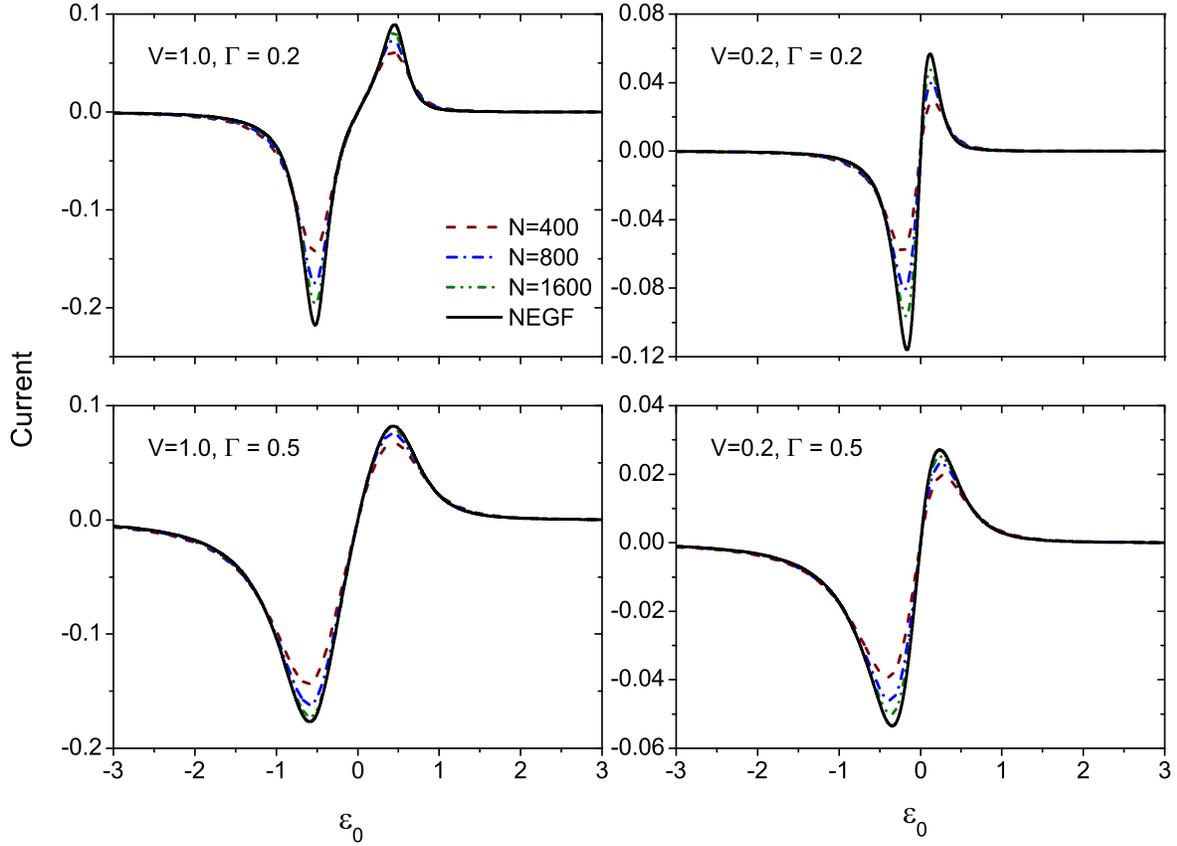}
\end{center}
\caption{The second-order perturbation theory correction to the current for the local Holstein model: Hartree term.
}
\label{hartree}
\end{figure}

\begin{figure}[t!]
\begin{center}
\includegraphics[width=1.\columnwidth]{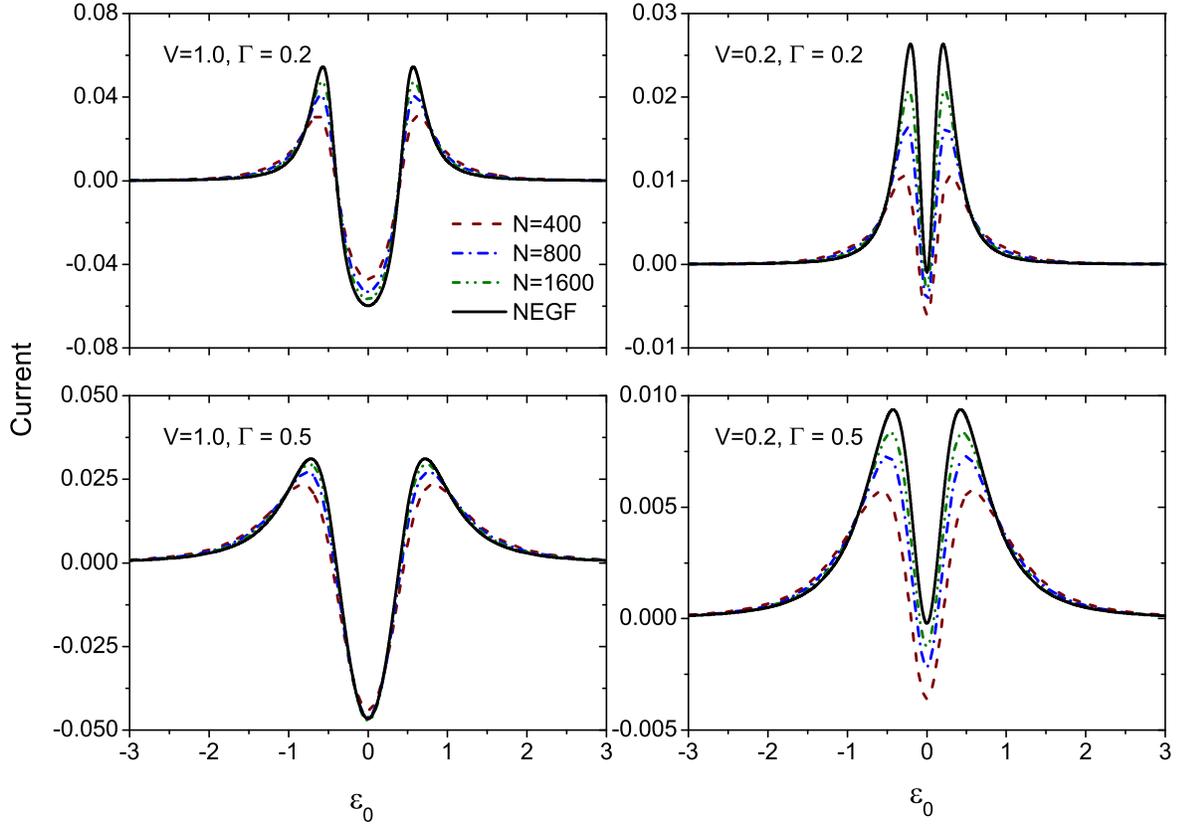}
\end{center}
\caption{
The second-order perturbation theory correction to the current for the local Holstein model: Fock term.
}
\label{fock}
\end{figure}

Let us now compare the results obtained with the present approach with those calculated  with the help of Keldysh NEGF.
For the case when the coupling to the left lead is proportional to the coupling to the right lead, the electron current through the quantum dot
can be computed directly from the retarded dot Green's  function, $G^r(\omega)$,
using the well known Meir-Wingreen formula~\cite{PhysRevLett.70.2601}. For the considered models, assuming that the
left and right leads are identical, $\Gamma_{L,R}(\omega)=0.5\Gamma(\omega)$, this formula takes the form
\begin{equation}
  J = \frac{s}{2\pi}\int d\omega [f_L(\omega) - f_R(\omega)] \Gamma(\omega)\mathrm{Im}G^r(\omega).
\end{equation}
Here $s$ is the spin degeneracy of the considered models: $s = 1$ for the model with electron-vibration coupling and $s = 2$ for the model with electron-electron interaction.
We will use the wide-band approximation for the electrode, so the imaginary part of the electrode self-energy, which is responsible for level broadening,
is energy independent, $\Gamma(\omega)=\Gamma$, while its real part vanishes.

The retarded Green's function is the solution of the Dyson equation
\begin{equation}
  G^r(\omega) = G^r_0(\omega) +  G^r_0(\omega)\Sigma^r(\omega) G^r(\omega),
\end{equation}
where $G^r_0(\omega) = (\omega - \varepsilon_0  + i\Gamma)^{-1}$ is the noninteracting retarded Green's function and $\Sigma^r(\omega)$ is retarded self-energy evaluated
in the presence of electron-electron or electron-vibration interaction.
Expanding   $\Sigma^r(\omega)$ with respect to electron-electron or electron-vibration coupling, $\Sigma^r(\omega)=\sum\limits_{p=1}\lambda^p\Sigma^r_p(\omega)$, we obtain  perturbative expansion of $G^r(\omega)$ and consequently of the current
\begin{align}\label{pe_NEGF}
& J = \frac{s}{2\pi}\int d\omega [f_L(\omega) - f_R(\omega)] \Gamma(\omega)
\notag\\
& \times\mathrm{Im}[G^r_0(\omega) +  \sum_{p=1} \lambda^pG^r_p(\omega)] =\sum_{p=0} \lambda^p J^{(p)}.
\end{align}
Here $J^{(0)}$ is the current through the system without interaction given by the standard Landauer formula.

In~\cite{dzhioev11a} we have shown that for the current through a system without interaction, $J^{(0)}$,  the exact agreement between NEGF and kinetic equation approach can
be achieved by increasing the density of states in the buffer zones. Below we show that  this is also true for correlated electronic systems.

In what follows, in the calculations based on the kinetic equation we will assume that $N$ single-particle levels in each buffer zone
are evenly distributed within the energy bandwidth  $[E_\mathrm{min};E_\mathrm{max}]=[-10,10]$.
This bandwidth is much larger than any energy parameter in the system, so it corresponds
to the  wide-band approximation used in NEGF calculations.  The tunneling coupling strength $t$ is computed from $\Gamma = 2\pi\eta t^2$, where
 $\eta = N/(E_\mathrm{max}-E_\mathrm{min})$ is density of states in the buffer zone.
 We note here that the main approximation in the derivation of the Lindblad master equation~\eqref{lindblad} is that
the single particle states in the buffer zone propagate in time as free states (\ref{sqrtn}). It is evident from Eq.(\ref{sqrtn})  that the larger the buffer zone, i.e. the larger the density of states $\eta$,  the better this approximation.
This will be also explicitly demonstrated in the numerical calculations below.
 The parameter $\gamma$ in the Lindblad operators is chosen to be $\gamma = 2\Delta\varepsilon$, where $\varepsilon$ is the energy spacing between the energy levels
in the buffer zone.  In addition, although it is not necessary, we use a symmetric applied voltage, $\mu_{L,R}=\pm0.5V$, and
 the temperature of the  electrodes is zero, $T=0$.

\begin{figure}[t!]
\begin{center}
\includegraphics[width=1.\columnwidth]{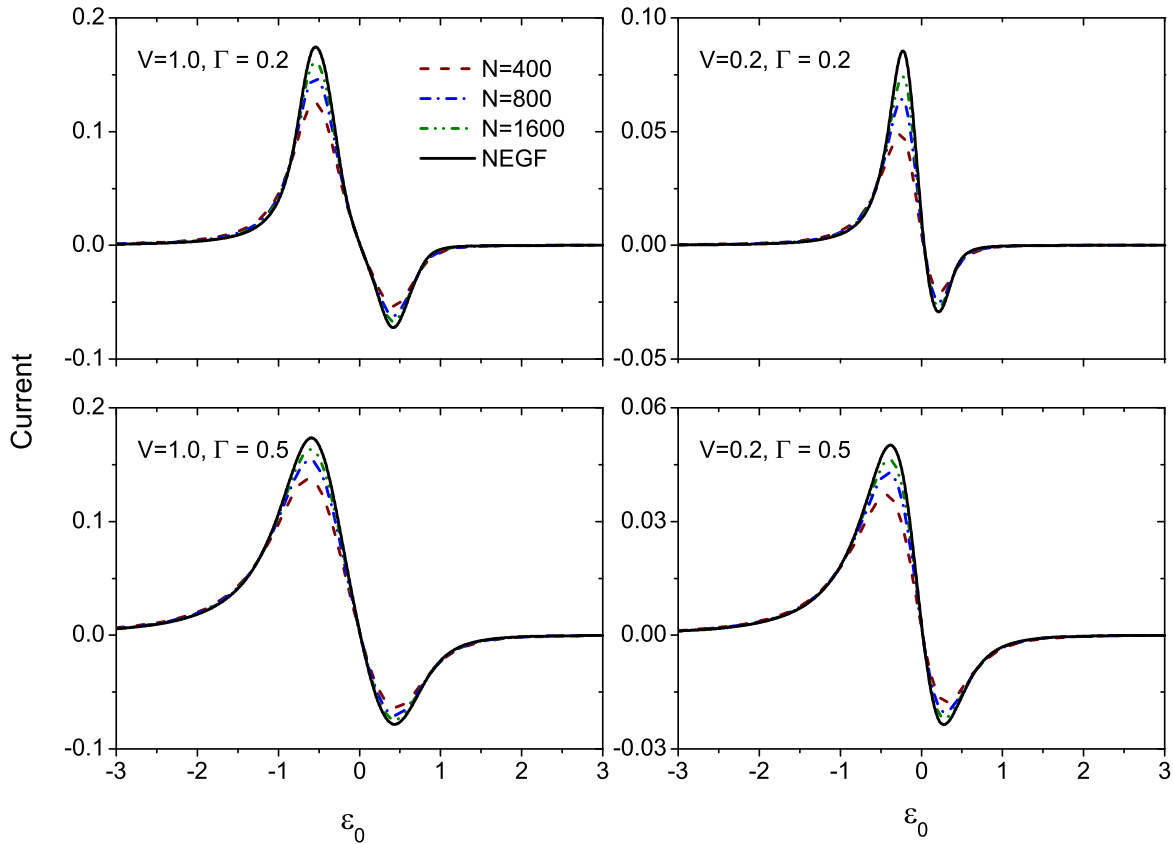}
\end{center}
\caption{The first-order perturbation theory correction to the current for the  Anderson model.}
\label{first_order_Hartree}
\end{figure}

\begin{figure}[t!]
\begin{center}
\includegraphics[width=1.\columnwidth]{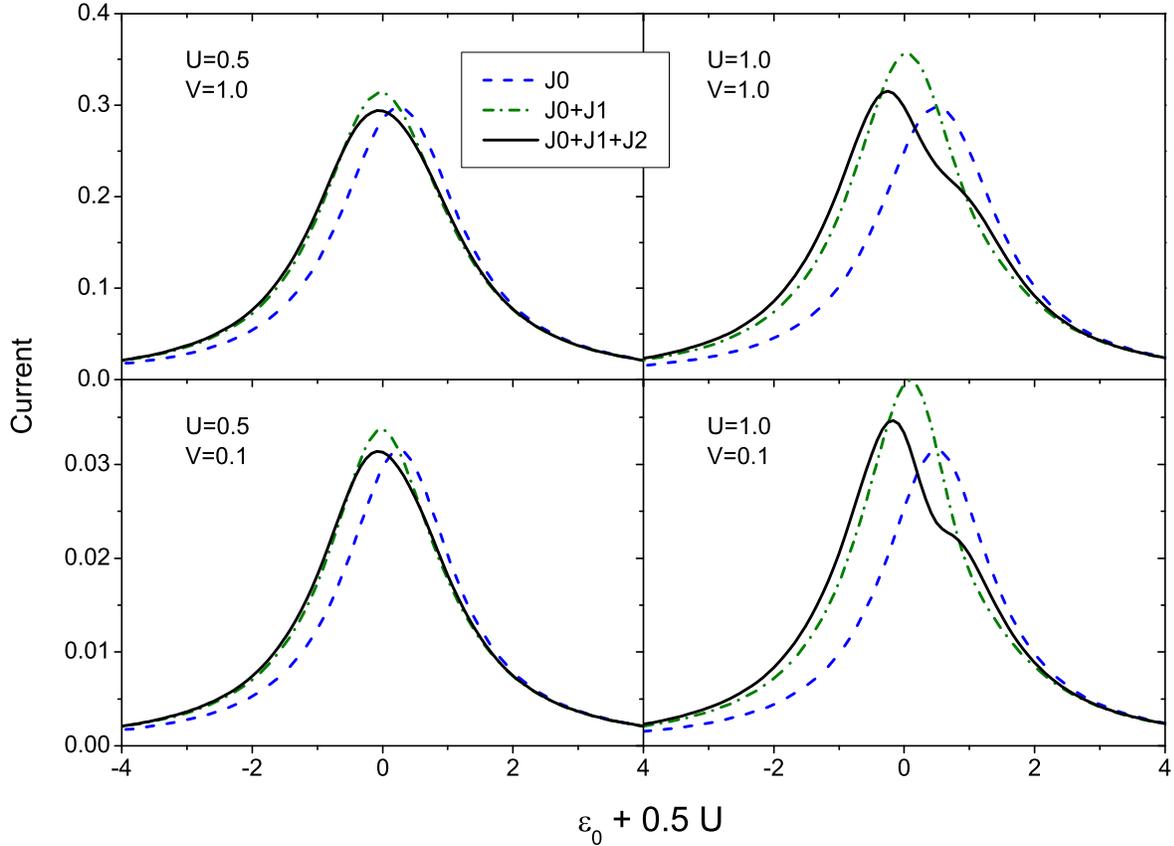}
\end{center}
\caption{The current through the Anderson model computed by taking into account the first- and second-order corrections.}
\label{total}
\end{figure}

At the beginning, we consider the system  with electron-vibronic interaction and compare the second-order correction to the current obtained in the section~\ref{EVC} with that calculated
using  NEGF formalism (\ref{pe_NEGF}). We use the following model parameters of the Hamiltonian~\eqref{H_s_vibr}: $\kappa=1.0$, $\omega_0=1.0$.

In NEGF formalism the second-order correction to the current arises from the retarded  self-energy  $\Sigma^r_2$ which contains contributions
from   Hartree and Fock diagrams, $\Sigma^r_2=\Sigma^r_H+\Sigma^r_F$. The Hartree self-energy is~\cite{Dash2010}
\begin{equation}
  \Sigma^r_H (\omega) = -  \frac{2\kappa^2}{\omega_0} n^{(0)},
\end{equation}
where $n^{(0)}$ is the electron level population in the zero-order approximation
\begin{equation}\label{pop_NEGF}
  n^{(0)}=\frac{\Gamma}{2\pi}\int d\omega \frac{f_L(\omega) + f_R(\omega)}{(\omega - \varepsilon_0 )^2 + \Gamma^2}.
\end{equation}
The expression for the Fock self-energy  is more complicated and  can be found elsewhere (see, for example, \cite{Egger2008}).

In Figs.~\ref{hartree} and \ref{fock}  we compare Hartree  and Fock   second-order  corrections to the current obtained within our approach with different size $N$ of buffer zone and
the exact ones. The corrections are shown as functions of the level energy, $\varepsilon_0$, for two values of the applied voltage $V$ and broadening $\Gamma$.
It is evident from the figures that the difference between exact and Lindblad equation based  results become negligible as we increase the leads density of states
in the buffer zone. The reason is that increasing the number of single-particle state in the buffer zones we make the approximation (iii), under which Lindblad master equation~(\ref{lindblad})
was derived, more justified.
The deviation of the results obtained  from the Lindblad kinetic equation and NEGF becomes smaller at the larger
applied voltage or $\Gamma$.

Now we compare first-order corrections to the current for the Anderson model.
We put $U=1.0$ for the strength of the Coulomb interaction.
Within the NEGF formalism the first order correction is solely due to Hartree  diagram and it is
\begin{align}
 J^{(1)} &= 4 \Gamma^2 U n^{(0)} \int\frac{d\omega}{2\pi} \frac{(f_L(\omega) - f_R(\omega))(\omega - \varepsilon_0)}{((\omega - \varepsilon_0 )^2 + \Gamma^2)^2},
\end{align}
where the population $n^{(0)}$ is given by~Eq.~\eqref{pop_NEGF}.

The results of numerical calculations
are shown in Fig.~\ref{first_order_Hartree} for different values of~$\Gamma$ and applied voltage~$V$.
As we can see  the results of the Lindblad equation approach converge to the exact results with increasing value of~$N$ and the convergence is faster
for larger values of applied voltage and $\Gamma$.

In Fig.~\ref{total} we show the current through the Anderson model computed by means of Lindblad equation by taking into account the first- and
second-order corrections. We take $N=1600$, so the obtained results correspond to NEGF ones. As we can see from the figure, the first- and second-order contributions
shift the maximum of the current towards the symmetric point $\varepsilon_0=-0.5U$.  The first-order correction increase the maximum current, while
the second-order correction acts in opposite  direction.
We also see from Fig.~\ref{total} that for a given $U$ the relative value of the first- and second-order corrections  show  little dependence on the applied voltage~$V$.
In contrast, in~\cite{dzhioev11b} we have observed that  nonequilibrium post-Hartree-Fock electronic correlations play
important role at larger applied voltages and, as a result, the second-order correction to the current become more pronounced  with increasing $V$.
This is due to the difference in the structure and spectrum of nonequilibrium quasiparticles. The quasiparticle spectrum, both $\psi$ and $\varphi$  amplitudes depend on the voltage in the
post-Hartree-Fock perturbation theory~\cite{dzhioev11b}, whereas in the present work the voltage enters only into  $\varphi$ amplitudes of the nonequilibrium quasiparticles through Fermi-Dirac occupation numbers of the buffer states.

\section{Conclusions}\label{conclusion}

We developed nonequilibrium many-body perturbation theory  for steady state density matrix and electric current through the region of interacting electrons. Our approach is based on the super-fermion representation of quantum kinetic equations.
We considered an quantum dot connected to the reservoir through the buffer zone (so-called embedded quantum dot).
The Lindblad type kinetic equation were obtained for the embedded quantum dot and the kinetic equation was converted to the non-Hermitian field theory in Liouville-Fock space via the tilde conjugation rules. The free-field state was defined as vacuum for non equilibrium quasiparticles and this state describes the  ballistic transport with the results  equivalent to the Landauer formulae.
We applied the nonequilibrium perturbation theory to compute corrections to nonequilibrium quasiparticle vacuum for  the system with electron-phonon and electron-electron correlations. The exact agreement with the Keldysh NEGF perturbation theory was observed for inelastic electron current through quantum dot.

\section*{References}



\begin{thebibliography}{10}

\bibitem{keldysh65}
L.~V. Keldysh.
\newblock Diagram technique for nonequilibrium processes.
\newblock {\em [Zh. Eksp. Teor. Fiz. 47, 1515 (1965)] Sov. Phys. JETP},
  20:1018, 1965.

\bibitem{imry1999}
Y.~Imry, R.~Landauer.
\newblock Conducrance viewed as transmission.
\newblock {\em Rev. Mod. Phys.}, 71(2):S306, 1999.

\bibitem{Caroli71}
C.~Caroli, R.~Combesco, P.~Nozieres, and D.~Saintjam.
\newblock Direct calculation of tunneling current.
\newblock {\em J. Phys. C}, 4(8):916, 1971.

\bibitem{galperin:035301}
M. Galperin, A. Nitzan, and M.~A. Ratner.
\newblock Inelastic effects in molecular junctions in the Coulomb and Kondo
  regimes: Nonequilibrium equation-of-motion approach.
\newblock {\em Phys. Rev. B}, 76(3):035301, 2007.

\bibitem{thoss11}
R.~H\"artle and M.~Thoss.
\newblock Vibrational instabilities in resonant electron transport through
  single-molecule junctions.
\newblock {\em Phys. Rev. B}, 83(12):125419, Mar 2011.

\bibitem{PhysRevB.69.245302}
A.~Mitra, I.~Aleiner, and A.~J. Millis.
\newblock Phonon effects in molecular transistors: Quantal and classical
  treatment.
\newblock {\em Phys. Rev. B}, 69(24):245302, Jun 2004.

\bibitem{dahnovsky:014104}
Y. Dahnovsky.
\newblock Ab initio electron propagators in molecules with strong
  electron-phonon interaction: II. Electron Green's function.
\newblock {\em J. Chem. Phys.}, 127(1):014104, 2007.

\bibitem{Dash2010}
L.~K. Dash, H.~Ness, and R.~W. Godby.
\newblock Nonequilibrium electronic structure of interacting single-molecule
  nanojunctions: Vertex corrections and polarization effects for the
  electron-vibron coupling.
\newblock {\em J. Chem. Phys.}, 132(10):104113, 2010.

\bibitem{PhysRevB.80.165305}
Yu. Dahnovsky.
\newblock Electron-electron correlations in molecular tunnel junctions: A
  diagrammatic approach.
\newblock {\em Phys. Rev. B}, 80(16):165305, 2009.

\bibitem{schmitt}
S. Schmitt and F.~B. Anders.
\newblock Comparison between scattering-states numerical renormalization group
  and the Kadanoff-Baym-Keldysh approach to quantum transport: Crossover from
  weak to strong correlations.
\newblock {\em Phys. Rev. B}, 81(16):165106, Apr 2010.

\bibitem{PhysRevB.75.075102}
P.~Darancet, A.~Ferretti, D.~Mayou, and V.~Olevano.
\newblock Ab initio $GW$ electron-electron interaction effects in quantum
  transport.
\newblock {\em Phys. Rev. B}, 75(7):075102, Feb 2007.

\bibitem{PhysRevB.77.115333}
K.~S.~Thygesen and A.~Rubio.
\newblock Conserving $GW$ scheme for nonequilibrium quantum transport in
  molecular contacts.
\newblock {\em Phys. Rev. B}, 77(11):115333, Mar 2008.

\bibitem{PhysRevB.79.155110}
C.~D.~Spataru, M.~S.~Hybertsen, S.~G.~Louie, and A.~J.~Millis.
\newblock $GW$ approach to Anderson model out of equilibrium: Coulomb blockade
  and false hysteresis in the $I-V$ characteristics.
\newblock {\em Phys. Rev. B}, 79(15):155110, Apr 2009.

\bibitem{thygesen07}
K.~S. Thygesen and A.~Rubio.
\newblock Nonequilibrium $GW$ approach to quantum transport in nano-scale
  contacts.
\newblock {\em J. Chem. Phys.}, 126:091101, 2007.

\bibitem{gurvitz96}
S.~A. Gurvitz and Ya.~S. Prager.
\newblock Microscopic derivation of rate equations for quantum transport.
\newblock {\em Phys. Rev. B}, 53(23):15932--15943, Jun 1996.

\bibitem{PhysRevB.78.235424}
M.~Leijnse and M.~R. Wegewijs.
\newblock Kinetic equations for transport through single-molecule transistors.
\newblock {\em Phys. Rev. B}, 78(23):235424, Dec 2008.

\bibitem{PhysRevB.74.235309}
U.~Harbola, M.~Esposito, and S.~Mukamel.
\newblock Quantum master equation for electron transport through quantum dots
  and single molecules.
\newblock {\em Phys. Rev. B}, 74(23):235309, Dec 2006.

\bibitem{PhysRevB.80.045309}
P.~Zedler, G.~Schaller, G.~Kiesslich, C.~Emary, and T.~ Brandes.
\newblock Weak-coupling approximations in non-Markovian transport.
\newblock {\em Phys. Rev. B}, 80(4):045309, Jul 2009.

\bibitem{PhysRevB.71.205304}
Xin-Qi Li, JunYan Luo, Yong-Gang Yang, Ping Cui, and YiJing Yan.
\newblock Quantum master-equation approach to quantum transport through
  mesoscopic systems.
\newblock {\em Phys. Rev. B}, 71(20):205304, May 2005.

\bibitem{PhysRevB.72.195330}
J.~N.~Pedersen and A.~Wacker.
\newblock Tunneling through nanosystems: Combining broadening with
  many-particle states.
\newblock {\em Phys. Rev. B}, 72(19):195330, Nov 2005.

\bibitem{ovchinnikov:024707}
I.~V. Ovchinnikov and D.~Neuhauser.
\newblock A Liouville equation for systems which exchange particles with
  reservoirs: Transport through a nanodevice.
\newblock {\em J. Chem. Phys.}, 122(2):024707, 2005.

\bibitem{dzhioev11a}
A.~A. Dzhioev and D.~S. Kosov.
\newblock Super-fermion representation of quantum kinetic equations for the
  electron transport problem.
\newblock {\em J. Chem. Phys.}, 134:044121, 2011.

\bibitem{dzhioev11b}
A.~A. Dzhioev and D.~S. Kosov.
\newblock Second-order post-Hartree--Fock perturbation theory for the electron
  current.
\newblock {\em J. Chem. Phys.}, 134:154107, 2011.

\bibitem{stability}
A.~A. Dzhioev and D.~S. Kosov.
\newblock Stability analysis of multiple nonequilibrium fixed points in
  self-consistent electron transport calculations.
\newblock {\em J. Chem. Phys.}, 135(17):174111, 2011.

\bibitem{schmutz78}
M.~Schmutz.
\newblock Real-time Green's functions in many body problems.
\newblock {\em Z. Physik. B}, 30:97 -- 106, 1978.

\bibitem{prosen08}
T.~Prosen.
\newblock Third quantization: a general method to solve master equations for
  quadratic open fermi systems.
\newblock {\em New Journal of Physics}, 10(4):043026, 2008.

\bibitem{Harbola2008}
U.~Harbola and S.~Mukamel.
\newblock Superoperator nonequilibrium Green's function theory of many-body
  systems; applications to charge transfer and transport in open junctions.
\newblock {\em Physics Reports}, 465(5):191 -- 222, 2008.

\bibitem{PhysRevLett.70.2601}
Y.~Meir, N.~S. Wingreen, and P.~A. Lee.
\newblock Low-temperature transport through a quantum dot: The Anderson model
  out of equilibrium.
\newblock {\em Phys. Rev. Lett.}, 70:2601--2604, Apr 1993.

\bibitem{Egger2008}
R.~Egger and A.~O. Gogolin.
\newblock Vibration-induced correction to the current through a single
  molecule.
\newblock {\em Phys. Rev. B}, 77:113405, Mar 2008.

\end{thebibliography}

\end{document}